\documentclass[twocolumn,showpacs,aps,prl,superscriptaddress]{revtex4}
\usepackage{graphicx,dcolumn,amsmath,epsfig,color,colordvi,amsfonts} 
\usepackage{fancyhdr,multirow,rotating,relsize} 
\RequirePackage{xspace}
\def\babar{\mbox{\slshape B\kern-0.1em{\smaller A}\kern-0.1em
    B\kern-0.1em{\smaller A\kern-0.2em R}}}
\def\CP    {\ensuremath{C\!P}\xspace}
\def\Dbar  {\kern 0.2em\overline{\kern -0.2em D}{}\xspace}
\def\Dz    {\ensuremath{D^0}\xspace}
\def\Dzb   {\ensuremath{\Dbar^0}\xspace}
\newcommand{\gevc}{\ensuremath{{\mathrm{\,Ge\kern -0.1em V\!/}c}}\xspace}
\newcommand{\mevcc}{\ensuremath{{\mathrm{\,Me\kern -0.1em V\!/}c^2}}\xspace}
\newcommand{\pppz}{\ensuremath{\pi^-\pi^+\pi^0}}
\newcommand{\kkpz}{\ensuremath{K^-K^+\pi^0}}
\newcommand{\CPV}{\ensuremath{CPV}}
\newcommand{\Dt}{\ensuremath{D}}
\newcommand{\thetaDcm}{\ensuremath{\theta_{\Dz}^{\mbox{\tiny{CM}}}}}
\long\def\inst#1{\par\nobreak\kern 4pt\nobreak
    {\it #1}\par\vskip 10pt plus 3pt minus 3pt}
\def\figurebox#1#2#3{
    \def\arg{#3}
    \ifx\arg\empty
    {\hfill\vbox{\hsize#2\hrule\hbox to 
	#2{\vrule\hfill\vbox to #1{\hsize#2\vfill}\vrule}\hrule}\hfill}
    \else {\hfill\epsfbox{#3}\hfill}
    \fi}

\begin{document}
\begin{flushleft}
\babar-PUB-07/074\\ 
SLAC-PUB-13058\\
arxiv:0802.4035 [hep-ex]\\
Phys. Rev. {\bf D78}, 051102(R) (2008)
\end{flushleft}

\title{{\large \bf \boldmath Search for \CP\ Violation in 
Neutral \Dt\ Meson Cabibbo-suppressed Three-body Decays}}
%% author list as of 03-Dec-2007 (546 authors)
%
\author{B.~Aubert}
\author{M.~Bona}
\author{Y.~Karyotakis}
\author{J.~P.~Lees}
\author{V.~Poireau}
\author{X.~Prudent}
\author{V.~Tisserand}
\author{A.~Zghiche}
\affiliation{Laboratoire de Physique des Particules, IN2P3/CNRS et Universit\'e de Savoie, F-74941 Annecy-Le-Vieux, France }
\author{J.~Garra~Tico}
\author{E.~Grauges}
\affiliation{Universitat de Barcelona, Facultat de Fisica, Departament ECM, E-08028 Barcelona, Spain }
% ...
\collaboration{The \babar\ Collaboration}
\noaffiliation

\date{\today}

\begin{abstract}
Using 385 fb${}^{-1}$ of $e^+e^-$ collision data collected at center-of-mass 
energies around 10.6 GeV, we search for time-integrated \CP 
violation in the Cabibbo-suppressed decays 
$\Dz/\Dzb \rightarrow \pi^-\pi^+\pi^0$ and 
$\Dz/\Dzb \rightarrow K^-K^+\pi^0$ with both 
model-independent and model-dependent methods. 
Measurements of the asymmetries in 
amplitudes of flavor states and \CP eigenstates 
provide constraints on theories beyond the standard model, 
some of which predict \CP violation in amplitudes 
at the 1\% level or higher.
We find no evidence of \CP violation and hence no conflict with 
the standard model.
 \end{abstract}
\maketitle
\begin{figure*}[!htbp]
\begin{tabular}{cc} 
\includegraphics[width=0.48\textwidth]{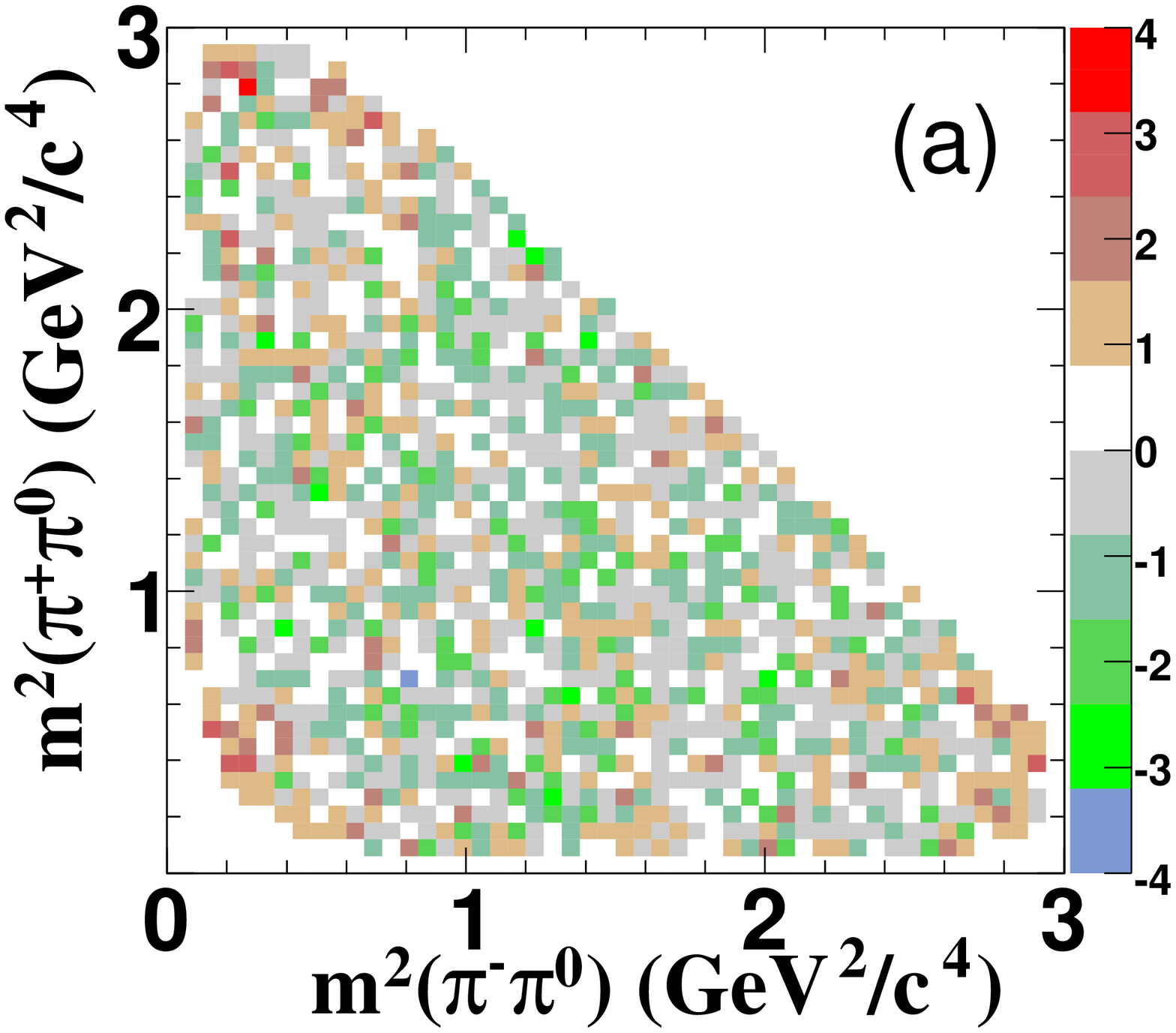} &
\includegraphics[width=0.48\textwidth]{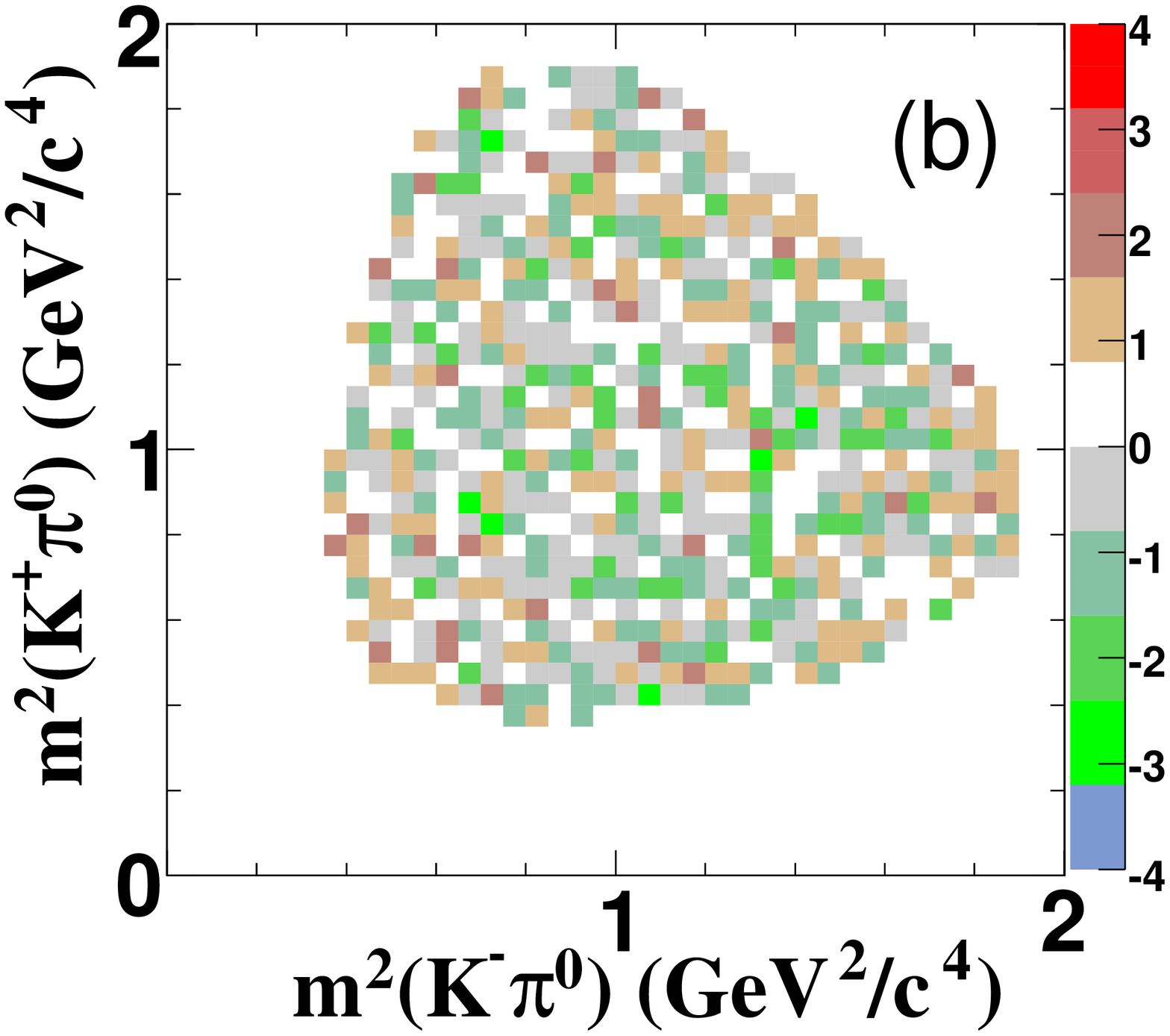}\\
\end{tabular}
\caption{ (color). Normalized residuals in Dalitz plot elements, 
defined in Eq.~\ref{Eq1},   
for (a) $\Dt\to\pppz$  and (b) $\Dt\to\kkpz$.} 
\label{Fig-1}
\end{figure*}
\begin{figure*}[!htbp]
\includegraphics[width=0.49\textwidth]{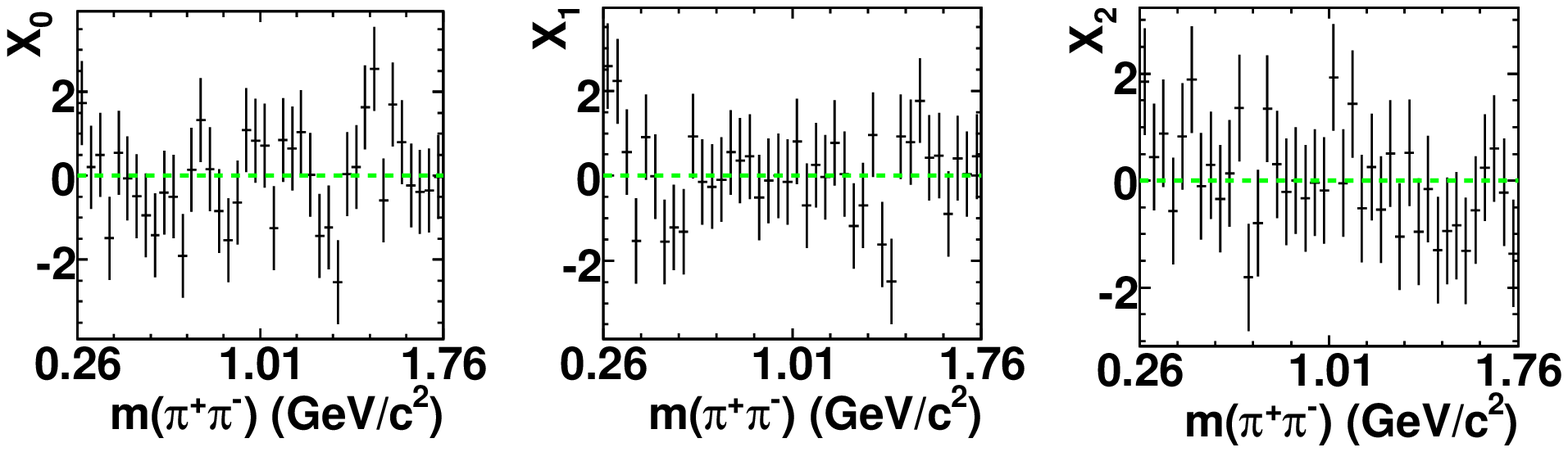}
\includegraphics[width=0.49\textwidth]{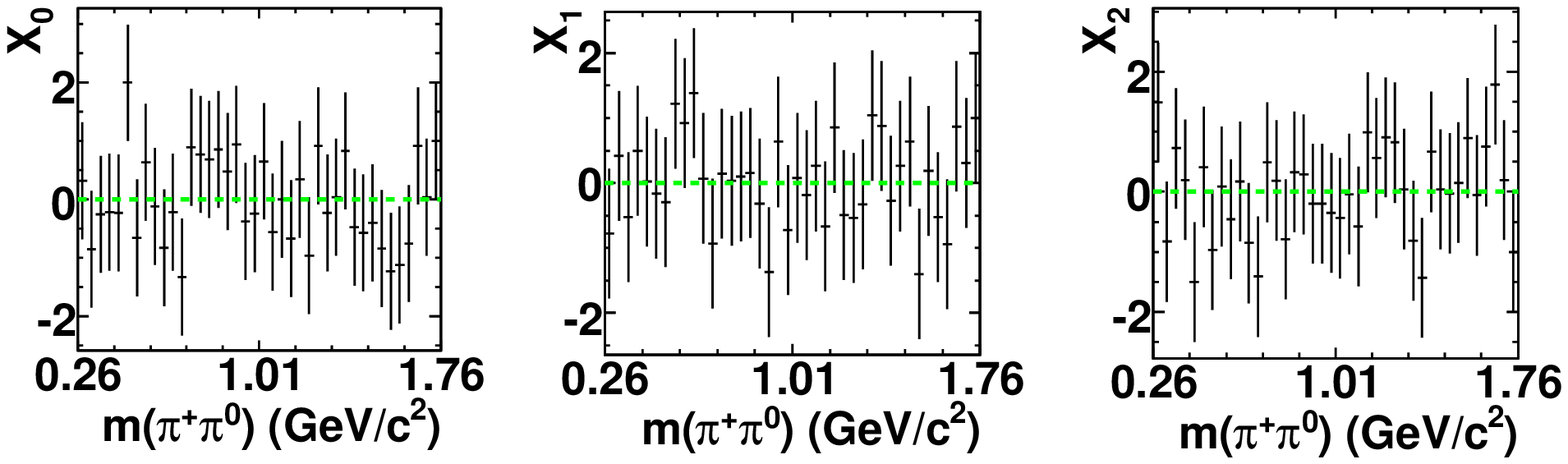}
\includegraphics[width=0.49\textwidth]{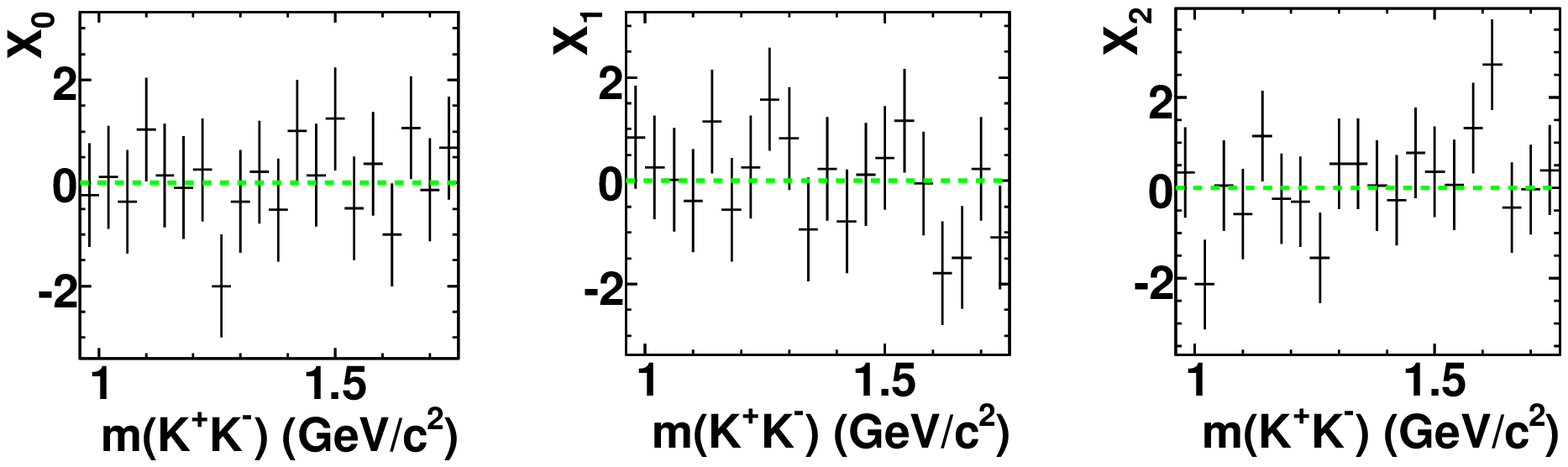}
\includegraphics[width=0.49\textwidth]{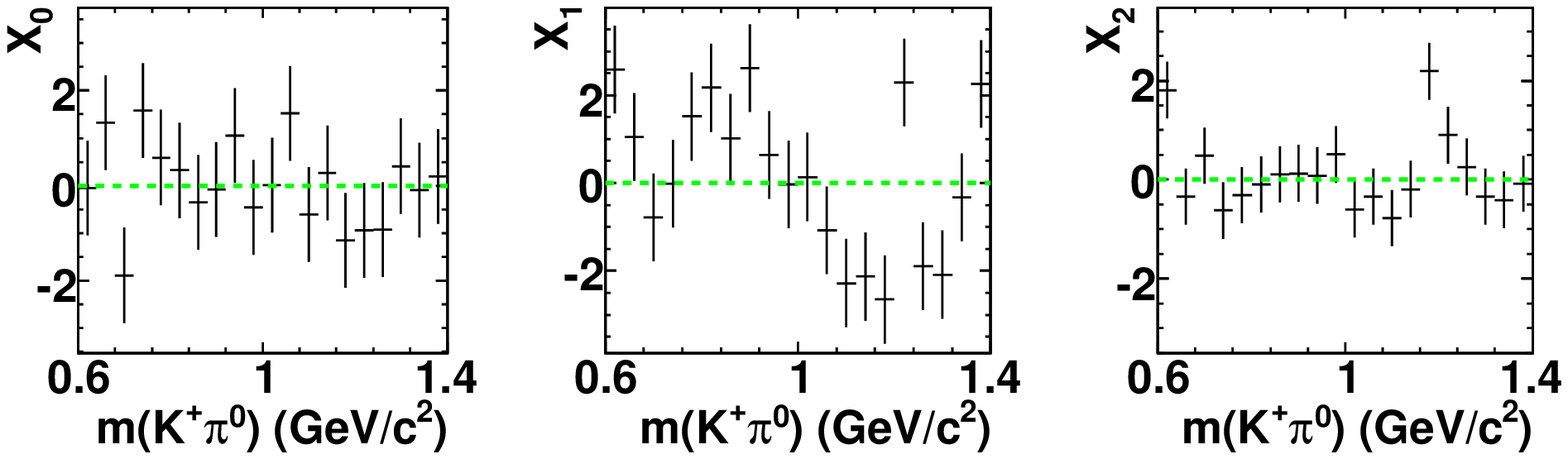}
\caption{(color online). Normalized residuals for the first 
three Legendre polynomial moments of the $\pi^-\pi^+$ (row 1),  
$\pi^+\pi^0$ (row 2), $K^-K^+$ (row 3), and  $K^+\pi^0$ (row 4) sub-systems. 
The confidence level for no \CP violation (dashed line) is obtained from 
the first eight moments. The error bars represent $\pm 1\sigma$.}
\label{Fig-2}
\end{figure*}

%%%%%%%%%%%%%%%%%%%%%
Charge-parity violation (\CPV)~\cite{cpv}, manifested as an asymmetry between 
the decay rates of a particle and its \CP-conjugate antiparticle, 
requires at least two interfering complex quantum mechanical amplitudes 
with different phases. 
The strong phase of each amplitude respects \CP\ symmetry while the 
weak phase changes sign under charge-conjugation.
In the standard model (SM), direct \CPV\ is due to relative weak phases 
that typically enter as a difference in phase between ``tree level'' 
and ``penguin''~\cite{penguin} SM amplitudes. 
The penguin amplitudes in charm decays are, however, too small 
(${\cal O}(0.1\%)$~\cite{kagan}) to provide significant \CPV.
Extensions of the SM introduce additional amplitudes of 
${\cal O}(1\%)$~\cite{kagan, bigi, petrov} with relative weak phases 
that can produce \CPV\ in charmed particle decays~\cite{pais}.
Current experimental 
searches~\cite{christian, coleman, belle, dcw, asner,cleoppp} 
are approaching this level of sensitivity.
Observation of \CPV\ with current experimental sensitivities 
would provide strong evidence of new physics.\\
%%%%%%%%%%%%%%%%%%%%%%%%%%%%%
%%%%%%%%%%%%%%%%%%%%%%%%%%%%%
\indent A recent theory paper~\cite{kagan} argues that singly 
Cabibbo-suppressed (SCS) \Dt\ (meaning either \Dz or \Dzb)
decays are uniquely sensitive to \CPV\ in 
$c\to u\bar d d, u\bar s s$ transitions and probe 
contributions from supersymmetric gluonic penguins. 
Such transitions do not affect the Cabibbo-favored ($c\to s \bar d u$) 
or doubly Cabibbo-suppressed ($c \to d \bar su$) decays.
Time-integrated \CP asymmetries in \Dt\ decays can have three components: 
direct \CPV\ in decays to specific states, indirect \CPV\ in 
\Dz--\Dzb mixing, and indirect \CPV\ in interference of decays with 
and without mixing. 
Indirect \CPV\ is predicted to be universal for amplitudes with 
final \CP eigenstates, but direct \CPV\ can be non-universal 
depending on the specifics of the new physics.\\ 
%%%%%%%%%%%%%%%%%%%%%%%%%%%%%
%%%%%%%%%%%%%%%%%%%%%%%%%%%%%
\indent We search for time-integrated \CPV\ in the three-body SCS decays 
$\Dt\rightarrow \pi^-\pi^+\pi^0, K^-K^+\pi^0$.
These decays proceed via \CP eigenstates (\textit{e.g.,} $\rho^0\pi^0$, 
$\phi\pi^0$) and also via flavor states (\textit{e.g.,} $\rho^{\pm}\pi^{\mp}$,
$K^{*\pm}K^{\mp}$), thus 
making it possible to probe \CPV\ in both types of amplitudes and 
in the interference between them. 
Measuring interference effects in a Dalitz plot (DP)
probes asymmetries in both the magnitudes and phases of the amplitudes, 
not simply in the overall decay rates.
We adopt four approaches in our search for evidence of \CPV,
three of which are model-independent.
First, we quantify differences between the \Dz 
and \Dzb DPs in two dimensions. 
Second, we look for differences in  
the angular moments of the \Dz and \Dzb intensity distributions.   
Third, in a model-dependent approach, we look for \CPV\ 
in the amplitudes describing intermediate states in the  
\Dz and \Dzb decays. 
Finally, we look for a phase-space-integrated asymmetry.
The first two methods are sensitive to differences in the shapes of
the \Dz and \Dzb DPs, allowing regions of
phase space with \CPV\ to be identified. 
The third method associates any \CPV\ observed using the first 
two methods with specific intermediate amplitudes. 
The last method is insensitive to differences in the DP 
shapes, so complements the other methods.
To minimize bias, we finalize the analysis procedure 
without looking at the data.\\
%%%%%%%%%%%%%%%%%%%%%%%%%
%%%%%%%%%%%%%%%%%%%%%%%%%
\indent We perform the present analysis using 385 fb${}^{-1}$ of $e^+e^-$ 
collision data collected at 10.58~GeV and 10.54~GeV center-of-mass (CM) 
energies with the \babar\ detector~\cite{detector} at the PEP-II storage 
rings. The event selection criteria are those used in our 
measurement of the branching ratios of the decays 
$\Dt\rightarrow \pppz$ and $\Dt\rightarrow \kkpz$~\cite{mybr}. 
In particular, we study \Dt\ mesons produced in 
$D^{*+} \rightarrow \Dz\pi^{+}$ and $D^{*-} \rightarrow \Dzb\pi^{-}$ 
decays that distinguish between \Dz and \Dzb. 
We require the \Dt\ candidate CM momentum $>2.77~\gevc$ and 
$|m_{D^{*\pm}} - m_{\Dt} - 145.4~\mevcc| < 0.6~\mevcc$. 
Here, $m$ refers to a reconstructed invariant mass. Around  
$\pm 1$ standard deviation of the nominal \Dt\ mass, we find  
$82468 \pm 321$ \pppz\  and $11278 \pm 110$ \kkpz\  signal events 
with purities of about 98\%.
We determine the signal reconstruction efficiency 
as a function of the position in the DP using 
simulated \Dz and \Dzb decays~\cite{mybr} from 
$e^+e^-\to c\overline c$ events, subjected 
to the same selection procedure that is applied to the data.\\
%%%%%%%%%%%%%%%%%%%%%%%%%%%%%%%%%%%%%%%%%%%%
%%%%%%%%%%%%%%%%%%%%%%%%%%%%%%%%%%%%%%%%%%%%
\indent A direct comparison of the efficiency-corrected and 
background-subtracted DPs for \Dz and \Dzb events is the 
simplest way to look for \CPV. Figure~\ref{Fig-1} shows the 
normalized residuals $\Delta$ in DP area elements, where 
\begin{equation}
\label{Eq1}
 {\Delta = {\left(n_{\Dzb} - R \cdot n_{\Dz}\right)/
{\sqrt{\sigma_{n_{\Dzb}}^2 + R^2 \cdot \sigma_{n_{\Dz}}^2}}}} \, ,
\end{equation}
and $n$ denotes the number of events in a DP element 
and $\sigma$ its uncertainty. 
The factor $R$, equal to $0.983 \pm 0.006$ for \pppz\ and $1.020 \pm 0.016$ 
for \kkpz, is the ratio of the number of efficiency-corrected 
\Dzb to \Dz events. 
This is introduced to allow for any asymmetry in the production cross 
section  due to higher order QED corrections or in the branching 
fractions for \Dz and \Dzb decay to the same final state.\\ 
%%%%%%%%%%%%%%%%%%%%%%%%%%%%%%%%%%%%%%%%%%%%%%%%%%%%%%%%%%%%%%%%%%%%%
%%%%%%%%%%%
%%%%%%%%%%%%%%%%%%%%%
\indent We calculate 
$\chi^2/\nu= (\sum_{i=1}^{\nu} \Delta_i^2)/\nu$, where 
$\nu$ is the number of DP elements: 1429 for \pppz\ and 726 for \kkpz.
In an ensemble of simulated experiments with no \CPV, we find 
the distribution of $\chi^2/\nu$ values to have a mean of 
1.012 $\pm$ 0.001 (1.021 $\pm$ 0.002) and an r.m.s. of 0.018 (0.036) 
for \pppz\ (\kkpz). 
The measured value in the data is 1.020 for \pppz\ 
and  1.056 for \kkpz, so we obtain a one-sided Gaussian 
confidence level (CL) for consistency with no \CPV\ of $32.8\%$ for 
\pppz\ and  $16.6\%$ for \kkpz. 
The same analysis procedure, when applied to
simulated samples with either 1\%  fractional change in magnitude  
or ${1}^\circ$  change in phase between the \Dz and \Dzb amplitudes 
for decay to any of the  
main resonant states, gives a $\chi^2/\nu$ 
that is about $2\sigma$ away from the no \CPV\ hypothesis.
Systematic  uncertainties  are small (as will be 
clear from the model-dependent results of 
Tables~\ref{tab:pipipi0DPFit}--\ref{tab:kkpi0DPFit}) and have not been 
included in the CL calculation.\\
%%%%%%%%%%%%%%%%%%%%%%%%%%%%%%%%%%%%%%%%%%%
\indent The angular moments of the cosine of the helicity angle 
of the \Dt\ decay products reflect the spin and mass structure of 
intermediate resonant and nonresonant amplitudes~\cite{mykkpi0}. 
We define the helicity angle $\theta_H$ for decays of the type 
$\Dt\to r(AB)~C$ as the angle between the momentum of $A$ in the 
$AB$ rest frame and the direction opposite to the \Dt\ momentum 
in that same frame. 
The angular moments~\cite{myhadron07} of order $l$ are defined as the 
efficiency-corrected invariant mass
distributions of events weighted by spherical harmonics 
$Y_l^0(\theta_H) = \sqrt{1 / {2\pi}}~ P_l(\cos\theta_H)$. 
Here $P_l$ are the Legendre polynomials of order $l$.
To study differences between the \Dz and \Dzb amplitudes,  
we calculate the quantities $X_l$ for $l=0-7$, where 
%%%%%%%
\begin{equation}
X_l  =  \frac{\left({\overline{P_l}} - R \cdot P_l\right)} 
   {\sqrt{\sigma_{\overline{P_l}}^2 + R^2 \cdot \sigma_{P_l}^2}} \, ,
\label{plerreqn}
\end{equation}
%%%%%%%
and $P_l$ ($\overline{P_l}$) are obtained from \Dz (\Dzb) events.
Higher moments are zero within errors in 
both data and simulation. For illustration, we show the $X_l$ 
distributions for $l=0-2$, in Fig.~\ref{Fig-2}.\\
%%%%%%%%%%%%%%
\indent We then define $\chi^2/\nu$ of the angular moment  
distributions of a two-body channel summed over all intervals 
in invariant mass as 
%%%%%%%%%%%
\begin{equation}
\chi^2/\nu=\left(\sum_{0}^k \sum_{i=0}^7 \sum_{j=0}^7 {X_i ~\rho_{ij} ~X_j} \right)/\nu,
\label{chi2formoments}
\end{equation}
%%%%%%%%%%% 
where $\nu=8k$, $k$ is the number of intervals, and 
$\rho_{ij}$  is the correlation coefficient between $X_i$, $X_j$:
%%%%%%%%%%%
\begin{equation}
\rho_{ij} = \frac{\left< X_i X_j \right> - 
  \left< X_i \right> \left< X_j \right>} 
{\sqrt{\left< X_i^2 \right> - \left< X_i \right>^2} \cdot  
  \sqrt{\left< X_j^2 \right> - \left< X_j \right>^2}}.
\label{correlationeqn}
\end{equation}
%%%%%%%%%%% 
We determine the $\rho_{ij}$ in each mass interval by simulating 
experiments with no \CPV. 
We test the method on real data by randomly assigning 
events as \Dz or \Dzb, and 
then calculating $\chi^2/\nu$ for the difference in their angular moments. 
We repeat this experiment 500 times and find the resulting $\chi^2/\nu$ 
distribution to be consistent with no \CPV, validating our 
calculation of $\rho_{ij}$.   
We then look at the \Dt\ flavor in the data and calculate the 
$\chi^2/\nu$ values for the two-body channels with charge combinations 
$+,-$ and $+,0$. Finally, we obtain a  one-sided Gaussian CL for consistency 
with no \CPV\ using the reference value and r.m.s. deviation from simulation.
We find the CL for no \CPV\ to be $28.2\%$ for the $\pi^+\pi^-$, 
$28.4\%$ for the $\pi^+\pi^0$, 
$63.1\%$ for the $K^+K^-$, and 
$23.8\%$ for the $K^+\pi^0$ sub-systems. 
Again, a 1\%  fractional change in magnitude 
or ${1}^\circ$  change in phase of any of the 
main resonant amplitudes gives a $\chi^2/\nu$ 
that is about $2\sigma$ away from the no \CPV\ hypothesis.\\
%%%%%%%%%%%%%%%%%%%%%%%%%%%%%
%%%%%%%%%%%
%%%%%%%%%%%%%%%%%%%%%%%%%%%%%%%%%%%%%%%%%%%
%%%%%%%%%%%%%%%%%%%%%%%%%%%%%
\indent The Dalitz plot amplitude $\cal{A}$ can be parametrized 
as a sum of amplitudes $A_r(s_+,s_-)$ for all relevant intermediate 
states $r$, each with a complex coefficient, i.e.,  
$\cal{A}$ = $\sum_r$ $a_r$ $e^{i\phi_r}$ $A_r(s_+,s_-)$, 
where $a_r$ and $\phi_r$ are real. 
Here $s_{+}$ and $s_{-}$ are the squared invariant masses of the 
pair of final state particles with charge combinations 
$+, 0$ and $-, 0$. 
The fit fraction for each process $r$ is defined as 
{\footnotesize{ $f_r \equiv \int \left|a_r A_r\right|^2 ds_+  ds_- / 
\int \left|\cal{A}\right|^2 ds_+  ds_-$}}.
We model incoherent, \CP-symmetric background 
empirically~\cite{mygamma,mykkpi0}. 
In the absence of \CPV, we expect the values of $a_r$ and 
$\phi_r$ (and hence $f_r$) to be identical for \Dz and \Dzb decay. 
The results obtained with this assumption 
are listed in Ref.~\cite{mygamma} for 
$\Dt\rightarrow \pppz$ and in Ref.~\cite{mykkpi0} for 
$\Dt\rightarrow \kkpz$. To allow the possibility of \CPV\  
in the present analysis, we let a second process -- not necessarily of 
SM origin -- contribute to each of the 
amplitudes $A_r$, thus permitting the $a_r$, $\phi_r$, $f_r$ 
for \Dz and \Dzb to differ. We summarize the results of the 
fit to the data in terms of the differences 
$\Delta{a_r} = a^{\Dzb}_r - a^{\Dz}_r$, 
$\Delta{\phi_r} = \phi^{\Dzb}_r - \phi^{\Dz}_r$, 
and $\Delta{f_r} = f^{\Dzb}_r - f^{\Dz}_r$ in 
Table~\ref{tab:pipipi0DPFit} for \pppz\ and in 
Table~\ref{tab:kkpi0DPFit} for \kkpz.
The \CP\ asymmetry in any amplitude, relative to that of the whole decay, 
is no larger than a few percent.\\
%%%%%%%%%%%%%%%%%%%%%%%%%
\begin{figure*}[!htbp]
\begin{center}
  \begin{tabular}{cc} 
    \includegraphics[width=0.45\textwidth]{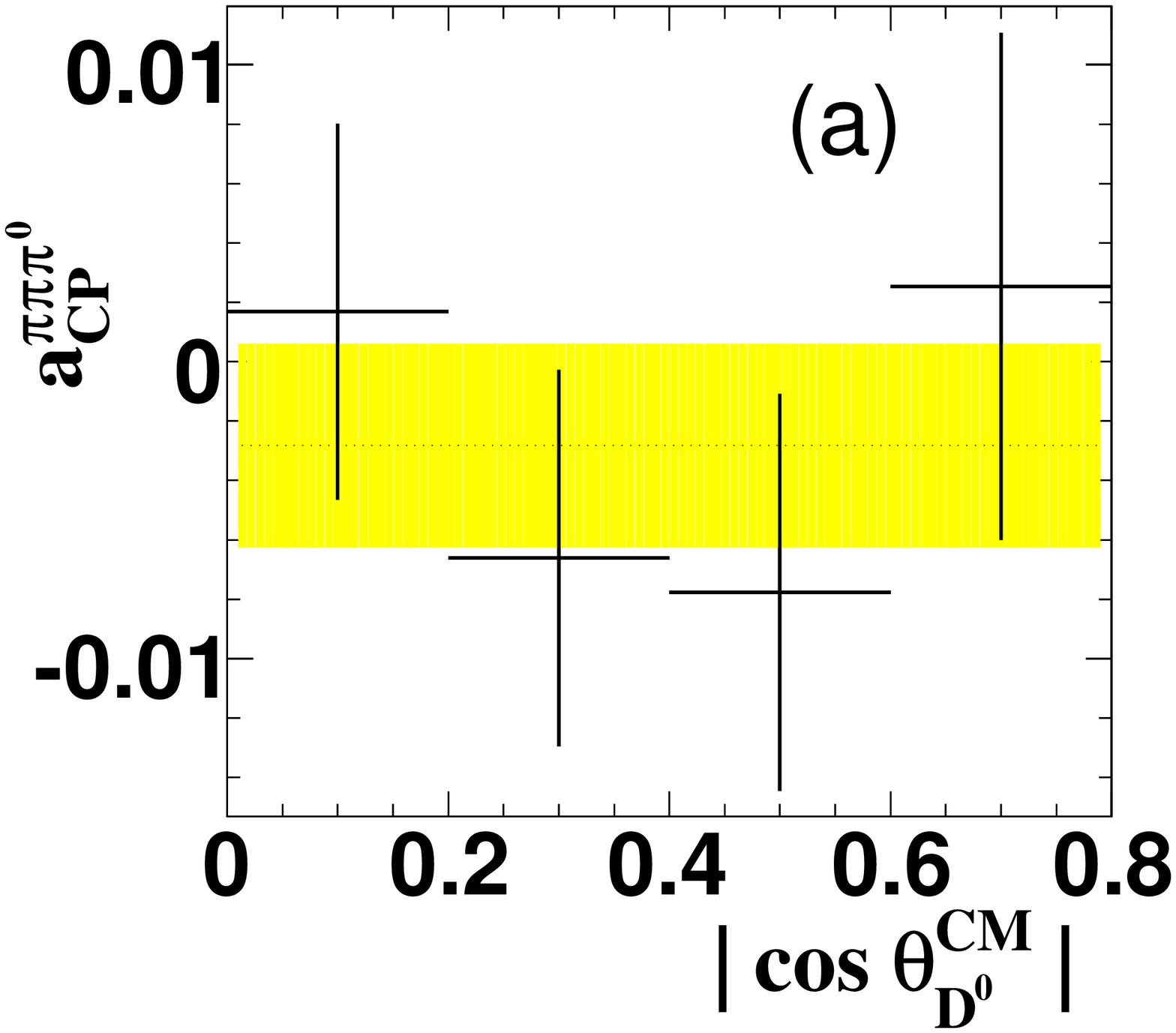} &
    \includegraphics[width=0.45\textwidth]{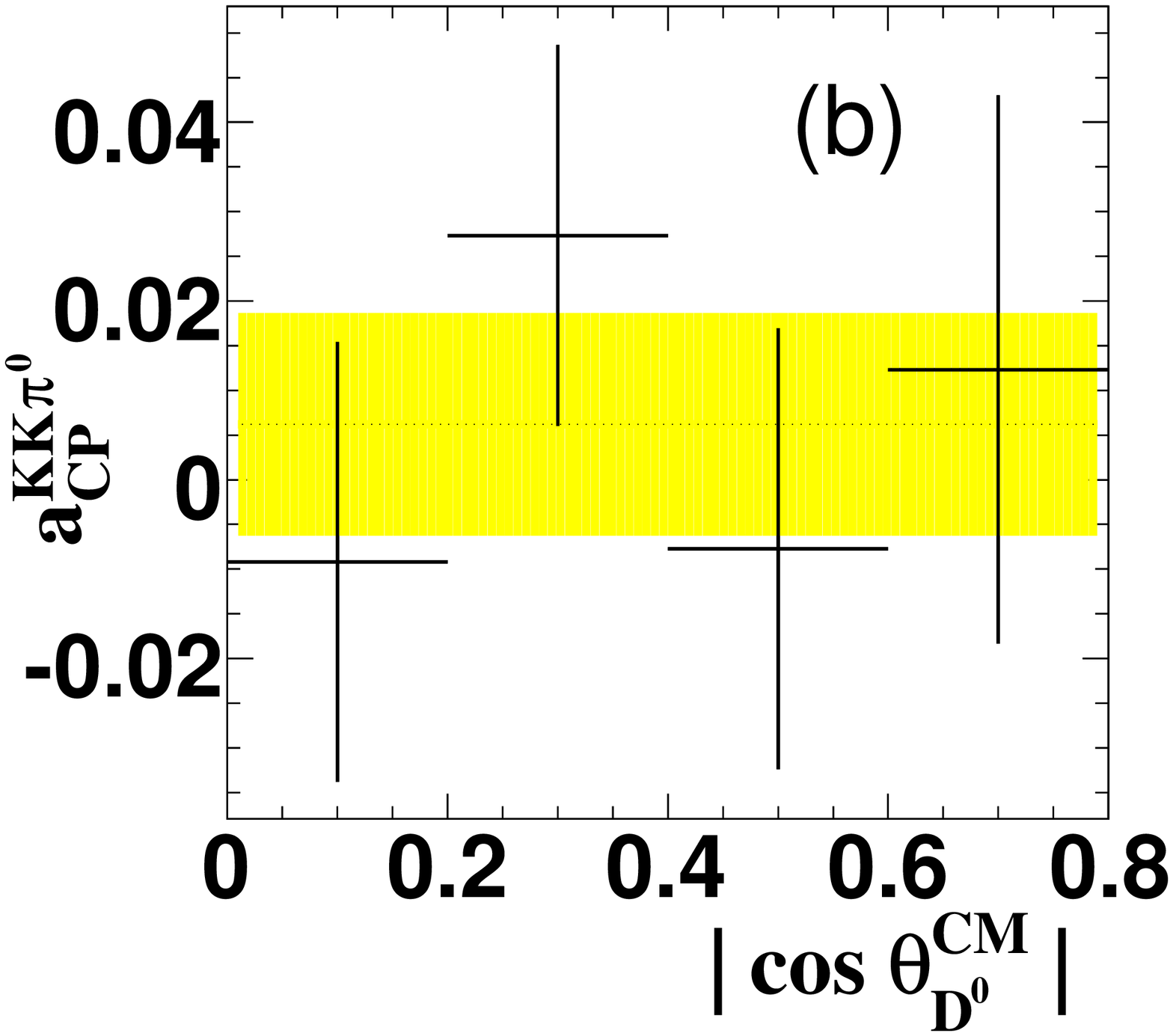} \\
  \end{tabular} 
\caption{(color online). Phase-space-integrated \CP asymmetry 
as a function of the cosine of the polar angle 
of the reconstructed \Dt\ candidate CM momentum for 
(a) $\Dt\to \pppz$ and (b) $\Dt\to \kkpz$ decays. 
The dashed lines represent the central values, and 
the shaded regions the 1$\sigma$ intervals.}
\label{Fig-3}
\end{center}
\end{figure*}
%%%%%%%%%%%%%%%%%%%%%%%%%%%%%%%%%  
%%%%%%%%%%%%%%%%%%%%%%%%
%%%%%%%%%%%%%%%%%%%%%%%%%
\begin{table*}[!htbp]
  \caption{\label{tab:pipipi0DPFit} Model-dependent \CP asymmetry in  
    the $\Dt\to \pi^- \pi^+ \pi^0$ Dalitz plots. 
    The first and second errors are statistical and systematic, 
    respectively. 
    For details on the Dalitz plot parametrization and the $a_r$, $\phi_r$,
    and $f_r$ values, see Ref.~\cite{mygamma}.
    As explained in text, $\Delta{f_r}$ is closely related to 
    $\Delta{a_r}$ and $\Delta{\phi_r}$}  
  \centering
  \begin{tabular}{lcccccccccccccccccccccccccccc}
    \hline
    \hline
    State &&&&&&& $f_r$ (\%) &&&&&&& $\Delta{a_r}$ (\%) &&&&&&& $\Delta{\phi_r}$ (${}^\circ$) 
    &&&&&&&  $\Delta{f_r}$ (\%) \cr
    \hline
    $\rho^+(770)$  &&&&&&& 68  &&&&&&&  -3.2$\pm$1.7$\pm$0.8  &&&&&&&  -0.8$\pm$1.0$\pm$1.0  &&&&&&&  -1.6$\pm$1.1$\pm$0.4\cr
    $\rho^0(770)$  &&&&&&& 26  &&&&&&&   2.1$\pm$0.9$\pm$0.5  &&&&&&&   0.8$\pm$1.0$\pm$0.4  &&&&&&&   1.6$\pm$1.4$\pm$0.6\cr
    $\rho^-(770)$  &&&&&&& 35  &&&&&&&   2.0$\pm$1.1$\pm$0.8  &&&&&&&  -0.6$\pm$0.9$\pm$0.4  &&&&&&&   0.7$\pm$1.1$\pm$0.5\cr
    $\rho^+(1450)$ &&&&&&& 0.1 &&&&&&&   2$\pm$11$\pm$8       &&&&&&&  -30$\pm$25$\pm$9      &&&&&&&   0.0$\pm$0.1$\pm$0.1\cr
    $\rho^0(1450)$ &&&&&&& 0.3 &&&&&&&   13$\pm$8$\pm$6       &&&&&&&  -1$\pm$14$\pm$3       &&&&&&&   0.1$\pm$0.2$\pm$0.1\cr
    $\rho^-(1450)$ &&&&&&& 1.8 &&&&&&&  -3$\pm$6$\pm$5        &&&&&&&   8$\pm$7$\pm$3        &&&&&&&  -0.2$\pm$0.3$\pm$0.1\cr
    $\rho^+(1700)$ &&&&&&& 4   &&&&&&&   19$\pm$27$\pm$9      &&&&&&&   9$\pm$7$\pm$3        &&&&&&&   0.4$\pm$1.0$\pm$0.4\cr
    $\rho^0(1700)$ &&&&&&& 5   &&&&&&&  -31$\pm$20$\pm$12     &&&&&&&  -7$\pm$6$\pm$2        &&&&&&&  -1.3$\pm$0.8$\pm$0.3\cr
    $\rho^-(1700)$ &&&&&&& 3   &&&&&&&  -3$\pm$14$\pm$11      &&&&&&&  -3$\pm$8$\pm$3        &&&&&&&  -0.5$\pm$0.6$\pm$0.3\cr
    $f_0(980)$     &&&&&&& 0.2 &&&&&&&   0.0$\pm$0.1$\pm$0.2  &&&&&&&  -3$\pm$7$\pm$4        &&&&&&&   0.0$\pm$0.1$\pm$0.1\cr
    $f_0(1370)$    &&&&&&& 0.4 &&&&&&&  -0.3$\pm$1.3$\pm$1.2  &&&&&&&   7$\pm$14$\pm$5       &&&&&&&  -0.2$\pm$0.1$\pm$0.1\cr
    $f_0(1500)$    &&&&&&& 0.4 &&&&&&&   0.4$\pm$1.1$\pm$0.7  &&&&&&&  -1$\pm$12$\pm$1       &&&&&&&   0.0$\pm$0.1$\pm$0.1\cr
    $f_0(1710)$    &&&&&&& 0.3 &&&&&&&  -3$\pm$3$\pm$2        &&&&&&&  -25$\pm$13$\pm$11     &&&&&&&   0.0$\pm$0.1$\pm$0.1\cr
    $f_2(1270)$    &&&&&&& 1.3 &&&&&&&   8$\pm$4$\pm$5        &&&&&&&   2$\pm$5$\pm$2        &&&&&&&   0.1$\pm$0.1$\pm$0.1\cr
    $\sigma(400)$  &&&&&&& 0.8 &&&&&&&  -0.3$\pm$0.7$\pm$2.0  &&&&&&&  -4$\pm$7$\pm$3        &&&&&&&  -0.1$\pm$0.1$\pm$0.1\cr
    Nonres         &&&&&&& 0.8 &&&&&&&   12$\pm$7$\pm$8       &&&&&&&   11$\pm$9$\pm$4       &&&&&&&   0.2$\pm$0.3$\pm$0.2\cr
    \hline
    \hline
  \end{tabular}
\end{table*}
%%%%%%%%%%%%%%%%%%%%%%%
%%%%%%%%%%%%%%%%%%%%%%%
\begin{table*}[!htbp]
\caption{Model-dependent \CP asymmetry in the $\Dt\to K^- K^+ \pi^0$ 
  Dalitz plots. The errors are statistical and systematic, 
  respectively. We show the $a_0(980)$ contribution, when it is included in 
  place of the $f_0(980)$, in square brackets.  For details 
    on the Dalitz plot parametrization and the $a_r$, $\phi_r$,
    and $f_r $ values, see Ref.~\cite{mykkpi0}. 
    We use Model-I of Ref.~\cite{mykkpi0} to obtain central values and 
    Model-II for study of systematic errors.}
\label{tab:kkpi0DPFit}
\begin{tabular}{lcccccccccccccccccccccccccccc}
  \hline
  \hline
    State &&&&&&& $f_r$ (\%) &&&&&&& $\Delta{a_r}$ (\%) &&&&&&& $\Delta{\phi_r}$ (${}^\circ$) 
    &&&&&&&  $\Delta{f_r}$ (\%) \cr
  \hline
  $K^*(892)^{+}$         &&&&&&& 45 &&&&&&&  2$\pm$3$\pm$2        &&&&&&&  10$\pm$12$\pm$3  &&&&&&&  0.8$\pm$1.1$\pm$0.4 \cr
  $K^*(1410)^{+}$        &&&&&&& 4  &&&&&&&  101$\pm$65$\pm$37    &&&&&&&  1$\pm$21$\pm$6   &&&&&&&  1.7$\pm$1.8$\pm$0.6 \cr
  $K^+\pi^0(\textit{S})$ &&&&&&& 16 &&&&&&& -130$\pm$64$\pm$51    &&&&&&& -9$\pm$10$\pm$6   &&&&&&& -2.3$\pm$4.7$\pm$1.0 \cr
  $\phi(1020)$           &&&&&&& 19 &&&&&&& -1$\pm$2$\pm$1        &&&&&&& -10$\pm$20$\pm$5  &&&&&&& -0.4$\pm$0.8$\pm$0.2 \cr
  $f_0(980)$             &&&&&&& 7  &&&&&&&  14$\pm$16$\pm$6      &&&&&&& -12$\pm$25$\pm$8  &&&&&&&  0.4$\pm$2.6$\pm$0.2 \cr
  $\left[a_0(980)^0\right]$&&&&&&&[6] &&&&&&& [19$\pm$16$\pm$6]  &&&&&&& [-7$\pm$16$\pm$8]&&&&&&&  [0.6$\pm$1.9$\pm$0.2] \cr
  $f_2'(1525)$           &&&&&&&0.1 &&&&&&& -38$\pm$74$\pm$8      &&&&&&&  6$\pm$36$\pm$12  &&&&&&&  0.0$\pm$0.1$\pm$0.3 \cr
  $K^*(892)^{-}$         &&&&&&& 16 &&&&&&&  1$\pm$3$\pm$1        &&&&&&& -7$\pm$4$\pm$2    &&&&&&&  1.7$\pm$1.3$\pm$0.4 \cr
  $K^*(1410)^{-}$        &&&&&&& 5 &&&&&&&  133$\pm$93$\pm$68     &&&&&&& -23$\pm$13$\pm$9  &&&&&&&  1.7$\pm$2.8$\pm$0.7 \cr
  $K^-\pi^0(\textit{S})$ &&&&&&& 3 &&&&&&&  8$\pm$68$\pm$36       &&&&&&& 32$\pm$39$\pm$14  &&&&&&&  0.4$\pm$2.4$\pm$0.5 \cr
  \hline 
  \hline
\end{tabular}
\end{table*}
%%%%%%%%%%%%%%%%%%%%%%%%%%%%%%
%%%%%%%%%%%%%%%%%%%%%%%%%%%%%%%%%  
\indent Systematic uncertainties in the quantities describing \CP asymmetries,
reported in Tables~\ref{tab:pipipi0DPFit}--\ref{tab:kkpi0DPFit},   
arise from experimental effects, and also from uncertainties in the 
models used to describe the data. 
We determine these separately, as described in 
Refs.~\cite{mygamma,mykkpi0}, and add them in quadrature. 
For all variations described below, we assign the maximum deviation 
from the central value as a systematic uncertainty, accounting for  
correlations among parameters. 
For resonance lineshapes and form-factors, we vary 
the parameters~\cite{pdg} by $\pm 1\sigma$.
Similarly, we vary the signal efficiency parameters for separately 
for \Dz and \Dzb 
events by $\pm 1\sigma$, the ratios of particle-identification 
rates in data and simulation by $\pm 1 \sigma$, and  
the background shapes by using simulation rather than data sidebands.  
We include uncertainties from \Dz--\Dzb misidentification, estimated from 
simulation, in the experimental systematic uncertainty.\\
%%%%%%%%%%%%%%%%%%%%%%%%%%%%%%%%%%%%%%%%%%%%%%%%%%%%%%%%%%%%%%%%%
%%%%%%%%%%%%%%%%%%%%%%%%%%%%%%%%%%%%%%%%%%%%%%%%%%%%%%%%%%%%%%%%%
\indent To this point, we have described the investigation of 
time-integrated \CP asymmetry in neutral $D$ meson decays using information  
from the DP distributions. 
Differences in the overall 
branching fractions for the \Dz and \Dzb decays to $\pi^-\pi^+\pi^0$, 
$K^-K^+\pi^0$ would also indicate time-integrated \CPV. 
This information is not captured by the differential comparisons 
of the DP structures already described, and is complementary to them. 
To correct for any production asymmetry in $D$-flavor assignment, 
we weight each event by the relative efficiency for flavor assignment, 
as described in Ref.~\cite{christian}. 
Since there is an asymmetry~\cite{christian} between the 
number of events reconstructed at forward and backward  
polar angles (\thetaDcm) of the \Dt\ candidate CM momentum,  
we extract the \CP asymmetry value,
$a_{\scriptscriptstyle \CP} \equiv 
{N_{\Dzb} - N_{\Dz} \over N_{\Dzb} + N_{\Dz}}$, 
in intervals of $|\cos{\thetaDcm}|$. 
Here, $N$ denotes the number of signal events.  
Any forward-backward asymmetry is
canceled by averaging over symmetric intervals in 
$\cos{\thetaDcm}$, as shown in Eqs.~3--5 of Ref.~\cite{christian}.
In Fig.~\ref{Fig-3} we show the $a_{\scriptscriptstyle \CP}$ for 
events in the \Dt\ mass window used in the DP analysis. 
We perform $\chi^2$ minimization to obtain the central values: 
[$-0.31$ $\pm$ $0.41$ (stat) $\pm$ $0.17$ (syst)] \% 
for $\pi^-\pi^+\pi^0$ and 
[$1.00$ $\pm$ $1.67$ (stat) $\pm$ $0.25$ (syst)] \% for $K^-K^+\pi^0$ 
final states. 
The systematic uncertainties result from 
signal efficiency, particle-identification, background treatment, and 
$\Dz-\Dzb$ misidentification. 
As a consistency check, we repeat the analysis with a 
larger \Dt\ mass window ($\pm 2.5\sigma$) and find consistent results:  
[$-0.28$ $\pm$ $0.34$ (stat) $\pm$ $0.19$ (syst)] \% for 
$\pi^-\pi^+\pi^0$ and 
[$0.62$ $\pm$ $1.24$ (stat) $\pm$ $0.28$ (syst)] \% for $K^-K^+\pi^0$.\\
%%%%%%%%%%%%%%%%%%%%%%%%
%%%%%%%%%%%%%%%%%%%%%%%%
\indent In summary, our model-independent and model-dependent analyses 
show no evidence of \CPV\ in the SCS decays $\Dt\to \pi^- \pi^+ \pi^0$
and $\Dt\to K^- K^+ \pi^0$. 
The intermediate amplitudes include well-defined flavor states 
(\textit{e.g.,} $\rho^{\pm}\pi^{\mp}$, $K^{*\pm}K^{\mp}$) and 
\CP-odd eigenstates (\textit{e.g.,} $\rho^0\pi^0$, $\phi\pi^0$). 
With the null results of Ref.~\cite{christian, coleman, belle, dcw} for  
\CP-even eigenstates $\Dt\to K^+K^-$ and $\Dt\to \pi^+\pi^-$, we conclude that 
any \CPV\ in the SCS charm decays occurs at a rate which is not larger than 
a few percent. 
These results are in accord with the SM predictions, and 
provide  constraints on some models beyond the SM~\cite{kagan}.\\
%%%%%%%%%%%%%%%%%%
%%%%%%%%%%%%%%%%%%
\indent We are grateful for the excellent luminosity and machine conditions
provided by our PEP-II colleagues, 
and for the substantial dedicated effort from
the computing organizations that support \babar.
The collaborating institutions wish to thank 
SLAC for its support and kind hospitality. 
This work is supported by
DOE
and NSF (USA),
NSERC (Canada),
CEA and
CNRS-IN2P3
(France),
BMBF and DFG
(Germany),
INFN (Italy),
FOM (The Netherlands),
NFR (Norway),
MES (Russia),
MEC (Spain), and
STFC (United Kingdom). 
Individuals have received support from 
the University Research Council (University of Cincinnati), 
the Marie Curie EIF (European Union), and
the A.~P.~Sloan Foundation.
%%%%%%%%%%%%%%%%%%%%%%%%


\begin{thebibliography}{99}
\bibitem{cpv}
J.H.~Christenson, J.W.~Cronin, V.L.~Fitch,  and  R.~Turlay,
Phys. Rev. Lett. {\bf 13}, 138 (1964). 
\bibitem{penguin}
M.A.~Shifman, ``{ITEP Lectures in Particle Physics}'', 
hep-ph/9510397, 5--6 (1995).
\bibitem{kagan} 
Y.~Grossman, A.L.~Kagan, and Y.~Nir, Phys. Rev. {\bf D75}, 036008 (2007).
\bibitem{bigi} 
S.~Bianco, F.L.~Fabbri, D.~Benson, and I.~Bigi, 
Riv. Nuovo Cim. {\bf 26N7}, 1 (2003).
\bibitem{petrov}
A.A.~Petrov, Phys. Rev. {\bf D69}, 111901 (2004).
\bibitem{pais}
A.~Pais and S.B.~Treiman, Phys. Rev. {\bf D12}, 2744 (1975).
Erratum: Phys. Rev. {\bf D16}, 2390 (1977).
\bibitem{christian}
B.~Aubert {\em et al.} (\babar\ Collaboration), 
Phys. Rev. Lett. {\bf 100}, 061803 (2008).
\bibitem{coleman}
B.~Aubert {\em et al.} (\babar\ Collaboration), arXiv:0712.2249
(to appear in Phys. Rev. {\bf D}) (2007).
\bibitem{dcw}
B.~Aubert {\em et al.} (\babar\ Collaboration), 
Phys. Rev. Lett. {\bf 91}, 121801 (2003).
\bibitem{belle}
M.~Staric {\em et al.} (Belle Collaboration), 
Phys. Rev. Lett. {\bf 98}, 211803 (2007).
\bibitem{asner}
D.~Asner {\em et al.} (CLEO Collaboration), 
Phys. Rev. {\bf D70}, 091101 (2004).
\bibitem{cleoppp}
D.~Cronin-Hennessy {\it et al.} (CLEO Collaboration), 
Phys.\ Rev.\ {\bf D72}, 031102 (2005).
Erratum: Phys.\ Rev.\ {\bf D75}, 119904 (2007).
\bibitem{detector} 
B.~Aubert {\em et al.} (\babar\ Collaboration), Nucl.\ Instr.\ and 
Methods {\bf A479}, 1 (2002).
\bibitem{mybr}
B.~Aubert {\em et al.} (\babar\ Collaboration), 
Phys. Rev. {\bf D74}, 091102 (2006).
\bibitem{mykkpi0}
B.~Aubert {\em et al.} (\babar\ Collaboration),
Phys. Rev. {\bf D76}, 011102 (2007). 
\bibitem{myhadron07}
K.~Mishra (\babar\ Collaboration), in \textit{proceedings 
of the XII International Conference on Hadron Spectroscopy}, 
edited by S.~Bianco, Frascati Physics Series Vol. 46 (
INFN Laboratori Nazionali di Frascati, Frascati, Italy, 2007),
p. 967.
\bibitem{mygamma} 
B.~Aubert {\em et al.} (\babar\ Collaboration), 
Phys. Rev. Lett. {\bf 99}, 251801 (2007).
\bibitem{pdg}
W.-M.~Yao {\em et al.} (PDG), 
J. Phys. ${\mathbf{G33}}$, 1 (2006). 
\end{thebibliography}
\end{document}